\documentclass[10pt,a4paper, DIV=14]{scrartcl}
\usepackage[utf8]{inputenc}
\usepackage[T1]{fontenc}
\usepackage[english]{babel}
\usepackage{textcomp}
\usepackage{amsmath,amssymb}
\usepackage{lmodern}
\usepackage{graphicx}
\usepackage[dvipsnames,svgnames]{xcolor}
\usepackage{microtype}
\usepackage{hyperref} \hypersetup{pdfstartview=XYZ}

\usepackage{hyperref}

\usepackage{caption}
\usepackage{multirow}
\usepackage{supertabular}

\usepackage{tikz}
\usetikzlibrary{patterns}

\usepackage{authblk}

\makeatletter
\newcounter {subsubsubsection}[subsubsection]
\renewcommand\thesubsubsubsection{\thesubsubsection .\@alph\c@subsubsubsection}
\newcommand\subsubsubsection{\@startsection{subsubsubsection}{4}{\z@}%
                                     {-3.25ex\@plus -1ex \@minus -.2ex}%
                                     {1.5ex \@plus .2ex}%
                                     {\normalfont\normalsize\bfseries}}
\newcommand*\l@subsubsubsection{\@dottedtocline{3}{10.0em}{4.1em}}
\newcommand*{\subsubsubsectionmark}[1]{}
\makeatother

\title{Modeling of dielectronic satellites to diagnose exotic states of matter created by XUV/X-ray free electron lasers - plasma ion electric microfield mixing dynamics rate (II): application to the $2l2l'$ doubly excited configuration of helium-like aluminium}

\author{Y.J. Aouad\footnote{PhD in Plasma Physics from UPMC "Université Pierre et Marie Curie" in the laboratory LULI "Laboratoire pour l'Utilisation des Lasers Intenses" - Ecole Polytechnique. Email: aouad852000@yahoo.fr}}
\affil{Theoretical physics, Atomic Physics in Plasmas}

\begin{document}

\maketitle

\begin{abstract}

In the present paper we give numerical estimations of the plasma ion electric microfield dependent rate introduced in \cite{Youcef7}. This rate was deduced from a quantum atomic density matrix formalism and corresponds to the mixing dynamics effect of energy levels by the plasma ion electric microfield. The rate in question is to be added to the usual collisional-radiative model for the modeling of dielectronic satellites originating from multi-excited atomic configurations to diagnose high density plasma regimes generated by the interaction of X-ray free electron lasers (XFEL's) with solid density matter. The obtained numerical values of this rate are compared to usual relaxation atomic rates of the collisional-radiative model in the case of three atomic energy levels of the doubly excited $2l2l'$ configuration of helium-like aluminium ($Z = 13$): $2p^2$ ${}^1D_2$, $2s2p$ ${}^1P_1$ and $2p^2$ ${}^1S_0$. The comparison is made for different values of the electronic density $n_e$ ($10^{+20}, 10^{+22}, 10^{+23}, 10^{+24} \ \text{cm}^{-3}$) and for the electronic temperature $T_e = 500 \ \text{eV}$. The numerical result shows that at high densities this rate is at the same order of magnitude as usual collisional-radiative rates and even exceed them in certain cases. This demonstrates the potential role of this rate for a better understanding of the heating mechanism underlying the evolution of a solid state matter to a plasma.
  
\end{abstract}

\tableofcontents
\bigskip

\section{Introduction}

\subsection*{X-ray emission originating from autoionizing and hollow ion configurations: time-resolved plasma spectroscopy}

\hspace{2mm} Spectroscopic methods based on dielectronic satellite lines originating from multi-excited and hollow ion atomic configurations play an important role in the study of the evolution of a solid state to warm dense matter (WDM) to strongly coupled plasma (SCP) in the context of matter irradiated by X-ray free electron lasers (XFEL's) (installations: LCLS 2011, XFEL 2011, SACLA XFEL 2011) \cite{Galtier1}. This is based on the properties of the XFEL’s itself, namely the creation of multi-excited states and hollow ion configurations by direct photoionization of the K and L atomic shells, as the energy per photon that constitutes the XFEL’s allows the ionization of the internal shells of atoms (ex: K and L) \cite{Rosmej5, Rosmej1}. These configurations are created principally in the dense plasma regime during the laser heating of matter. 
\\
\\
Multi-excited configurations are autoionizing ones and are characterized by very short lifetimes of about $1-10$ $\text{fs}$ due to their high autoionization rates ($\Gamma \approx 10^{+13} - 10^{+16}$ s$^{-1}$). The short lifetime leads likewise to an intrinsic time resolution for the corresponding dielectronic satellite emission. The high intensity of the XFEL’s laser causes high number of photoionization events of the internal atomic shells and high population of multi-excited states and hollow ion configurations. That is why the development of new spectroscopic methods based on the multi-excited and hollow ions atomic configurations is enormously advantageous for the study of the coupled plasma regimes. 
\\
\\
The use of dielectronic satellites for the plasma diagnostic propose is based on the analysis of the spectral distribution $I_{satellite}(\omega)$ (energy emitted per unit time, per unit area, per unit solid angle and unit angular frequency, $[\text{eV} \ \text{s}^{-1} \ \text{cm}^{-2} \ \text{sr} \ 2\pi \text{Hz}]$) given by:
\\
\begin{equation}\label{I_omega}
I_{satellite}(\omega) = \frac{1}{4\pi} \sum_{\gamma J \rightarrow \gamma' J'} \hbar \omega_{\gamma J \rightarrow \gamma' J'} \ N_{\gamma J} \ A_{\gamma J \rightarrow \gamma' J'} \ \Phi_{\gamma J \rightarrow \gamma' J'}(\omega)
\end{equation}
\\
where in Eq.\ref{I_omega}, $\hbar \omega_{\gamma J \rightarrow \gamma' J'}$ is the energy corresponding to the transition between the two energy levels $\gamma J$ and $\gamma' J'$ ($J$ is the total kinetic momentum and $\gamma$ account for additional quantum numbers), $N_{\gamma J} $ is the population of the upper level $\gamma J$ belonging to a multi-excited configuration, $A_{\gamma J \rightarrow \gamma' J'}$ is the Einstein radiative decay rate for the transition $\gamma J \rightarrow \gamma' J'$ and $ \Phi_{\gamma J \rightarrow \gamma' J'}$ is the line profile related to the broadening of the satellite line $\gamma J \rightarrow \gamma' J'$ due to different processes, e.g., finite life-time of the energy level $\gamma J$, Doppler broadening, Stark broadening.
\\
\\
As seen in Eq.\ref{I_omega}, the modeling of the spectral distribution $I_{satellite}(\omega)$ requires the calculation of populations $N_{\gamma J}$ of the emitting energy levels of multi-excited configurations. These populations are solution of a system of equations that couple all ground, excited and multiexcited energy levels considered in the studied system. The usual method consists of solving the so-called collsional-radiative kinetic system (CR) which takes into account different atomic transfer processes in a plasma. The time evolution of the atomic energy level Populations in the usual CR model is given by the following system of differential equations:
\\
\begin{equation}\label{Diff_CR_Pop}
\frac{dN_{\gamma J}}{dt} = -N_{\gamma J} \sum_{\gamma' J'} W_{\gamma J, \gamma' J'} + \sum_{\gamma' J'} N_{\gamma' J'} W_{\gamma' J', \gamma J}
\end{equation}
\\
in Eq.\ref{Diff_CR_Pop}, $W_{\gamma J, \gamma' J'}$ is the collisional-radiative rate from the energy level $\gamma J$ to the energy level $\gamma' J'$ and it takes into account all transfer processes in a plasma like, for instance, spontaneous radiative decay, electron collisional excitation and dexciatation, autoionization, three body recombination, ionization by electrons, photo-ionization, photo-recombination and so on:
\\
\begin{equation}\label{W_CR}
 W_{\gamma J, \gamma' J'} = \sum_{P \in \{Processes\}} [W_{\gamma' J', \gamma J}]^{P}
\end{equation}
\\
In the stationary regime, the system of equations Eq.\ref{Diff_CR_Pop} transforms to:
\\
\begin{equation}\label{Diff_CR_Pop_Stat}
-N_{\gamma J} \sum_{\gamma' J'} W_{\gamma J, \gamma' J'} + \sum_{\gamma' J'} N_{\gamma' J'} W_{\gamma' J', \gamma J} = 0
\end{equation}
\\
The use of Eqs.\ref{I_omega},\ref{Diff_CR_Pop_Stat} to diagnose dense plasma regimes is subject to a more investigation as in this regime the plasma ion electric microfield mixing dynamics has to be taken into account in the calculation of atomic energy level populations \cite{Youcef7}. This has also a consequence in the calculation of spectral line shapes \cite{Youcef6}. It has been proposed \cite{Youcef7} to rewrite the system of equations Eq.\ref{Diff_CR_Pop_Stat} by adding a new rate taking into account the plasma ion electric microfield mixing dynamics and to consider this new rate on the same footing as the collisional-radiative rates ($W_{\gamma J, \gamma' J'}$). This is due to the fact that for autoionizing configurations (multi-excited and autoionizing configurations), electron densities justifying the Boltzmann distribution of energy levels inside a configuration exceed solid density. Because autoionizing configurations are characterized by a high autoionization rates \cite{Rosmej1}. At the same time, at high density regimes non-local thermodynamic equilibrium system of equations on atomic energy level populations must take into account the plasma ion electric microfield as it will be shown in this paper that the new rate in question is of the same order of magnitude as the usual collisional-radiative rates and even it exceed them in certain cases. 


\section{Plasma ion electric microfield dependent atomic energy level population kinetics}
\hspace{2mm} It has been proposed in \cite{Youcef7} to rewrite the usual stationary CR model as follows: 
\\
\begin{equation}\label{Stat_N_gammaJ}
-N_{\gamma J} \times \sum_{\gamma' J'} \left[  W_{\gamma J,\ \gamma' J'} + W_{\gamma J,\ \gamma' J'}(E) \right] + \sum_{\gamma' J'} N_{\gamma' J'} \times \left[ W_{\gamma' J',\ \gamma J} + W_{\gamma' J',\ \gamma J}(E)\right] = 0
\end{equation}
\\
where in Eq.\ref{Stat_N_gammaJ}, $E$ is the amplitude of the plasma ion electric microfield and the new rate $W_{\gamma J,\ \gamma' J'}(E)$ related to the mixing dynamics effect is given by:
\\
\begin{multline}\label{W_E_gammaJ_gammaJ}
W_{\gamma J,\ \gamma' J'}(E) =\frac{C^{\textbf{J'11},\ 00}_{J}}{g_{\gamma J}}  \times  \left(\frac{e a_{0} E}{\hbar}\right)^{2} \times \left(-1\right)^{J-J'} \times \mid< \gamma J \mid\mid \textbf{\emph{P}}^{\textbf{(1)}} \mid\mid \gamma' J' >\mid^{2} \times \\
\frac{\left[ \sum_{\gamma''J''} W_{\gamma J,\ \gamma'' J''} +  \sum_{\gamma''J''} W_{\gamma' J',\ \gamma''J''} \right]}{\left[ \frac{E_{\gamma J}-E_{\gamma' J'}}{\hbar} \right]^2 +\frac{1}{4}\left[ \sum_{\gamma'' J''} W_{\gamma J,\ \gamma'' J''} +  \sum_{\gamma''J''} W_{\gamma' J',\ \gamma'' J''} \right]^2}
\end{multline}
\\
where in Eq.\ref{W_E_gammaJ_gammaJ}, $e$ is the absolute value of the electronic charge, $a_0$ is the Bohr radius ($\simeq 0.529 \times 10^{-8} \text{cm}$), $\hbar$ is the reduced Planck constant, $g_{\gamma J}$ is the statistical weight of the energy level $\gamma J$ ($g_{\gamma J} = 2J+1$), $< \gamma J \mid\mid \textbf{\emph{P}}^{\textbf{(1)}} \mid\mid \gamma ' J' >$ is the reduced matrix element of the dipole moment  $\textbf{\emph{P}}^{\textbf{(1)}}$ of the emitter in units of $ea_0$ \cite{Cowan1}, $W_{\gamma J, \gamma' J'}$ is the usual collisional-radiative rate, $E_{\gamma J}$ is the energy of the energy level $\gamma J$ and the coefficient $C^{\textbf{J'11},\ 00}_{J}$ has been defined in \cite{Youcef7} expressed in terms of sums involving 3-$j$ symbols and is given by:
\\
\begin{equation}\label{C1100_gammaJ}
C^{\textbf{J11},\ 00}_{J'} \equiv \sum_{M,\ M',\ M''} (-1)^{J-M} (-1)^{J'-M'} \left(\begin{array}{clcr}
J & \ \ J' & 1\\
M & -M' & 0  \end{array}\right) \times \left(\begin{array}{clcr}
J & \ \ J' & 1\\
M & -M'' & 0  \end{array}\right)
\end{equation}
\\
In Eq.\ref{C1100_gammaJ}, the coefficient $C^{\textbf{J11},\ 00}_{J'}$ is symmetric to the respect to the permutation of $J$ and $J'$: 
\\
\begin{equation}\label{Symmetri_C1100_gammaJ}
C^{\textbf{J11},\ 00}_{J'} = C^{\textbf{J'11},\ 00}_{J}
\end{equation}
\\
and is equal to:
\\
\begin{equation}\label{C1100_gammaJ_delta}
C^{\textbf{J11},\ 00}_{J'} = C^{\textbf{J'11},\ 00}_{J} = \delta(J1J') \times \frac{\left(-1\right)^{J-J'} }{3}
\end{equation}
\\
In Eq.\ref{C1100_gammaJ_delta}, $\delta(J1J')$ expresses the triangular relation condition for $J$, $1$ and $J'$ ($\delta(J1J')=+1$ when $J$, $1$ and $J'$ verify the triangular relation and $\delta(J1J')=0$ otherwise). It is to note that the coefficient $C^{\textbf{J'11},\ 00}_{J}$ provides a selection rule for the rate $W_{\gamma J,\ \gamma' J'}(E)$ to the respect to the two total kinetic momentums $J$ and $J'$ by $\delta(J1J')$. And from Eqs.\ref{W_E_gammaJ_gammaJ},\ref{Symmetri_C1100_gammaJ}, one has to note the following ratio:
\\
\begin{equation}\label{Ratio_W_E_gammaJ_gammaJ_E}
\frac{W_{\gamma J,\ \gamma' J'}(E)} {W_{\gamma' J',\ \gamma J}(E)}= \frac{g_{\gamma' J'}}{g_{\gamma J}}
\end{equation}
\\
It is to note that the solution $N_{\gamma J}$ of the system of equations Eq.\ref{Stat_N_gammaJ} is field dependent: 
\\
\begin{equation}\label{N_E}
N_{\gamma J} \equiv N_{\gamma J}(E)
\end{equation}
\\
The final result has to be averaged over the stationary plasma ion electric microfield distribution function $P(E)$:
\\
\begin{equation}\label{N_average}
<N_{\gamma J}> = \int dE \ P(E) \ N_{\gamma J}(E)
\end{equation}


\section{Numerical application to the doubly excited 2l2l' configuration of helium-like aluminium}
\hspace{2mm} In order to evaluate and compare the values of the plasma ion electric microfield dependent rate Eq.\ref{W_E_gammaJ_gammaJ} to usual radiative-collisional rates, we apply Eq.\ref{W_E_gammaJ_gammaJ} to three doubly excited atomic energy levels of helium-like aluminium ($Z = 13$): $2p^2$ ${}^1D_2$, $2s2p$ ${}^1P_1$ and $2p^2$ ${}^1S_0$. These three energy levels are populated by dielectronic capture ($DC$) from the hydrogen-like aluminium energy level $1s$. The atomic structure characteristics, .i.e., energies, statistical weights of these energy levels and collisional-radiative relaxation constants, are obtained by the use of the relativistic atomic code $\text{FAC}$ (Flexible Atomic Code) \cite{Gu1}. The energy levels $2p^2$ ${}^1D_2$, $2s2p$ ${}^1P_1$ and $2p^2$ ${}^1S_0$ are depopulated by the autoionization process ($\Gamma$) to the level $1s$ and by the micro-reversibility principle, they are populated by the dielectronic capture ($DC$) from this level. Collisional electronic inelastic relaxation rates $C$ between these $2l2l'$ energy levels is considered and calculated with the code $\text{FAC}$ \cite{Gu1}.The plasma ion static electric microfield $\vec{E}$ couple the three $2l2l'$ energy levels. Relaxations of these $2l2l'$ energy levels to the ground $1s^2$ and singly-excited $1s2s$ and $1s2p$ helium-like aluminium configurations are introduced by the total rates $\gamma(2p^2 \ {}^1D_2)$, $\gamma(2s2p \ {}^1P_1)$ and $\gamma(2p^2 \ {}^1S_0)$ that include radiative parts $\gamma^r$ ($\equiv A$) and collisional parts $\gamma^C$ ($\equiv C$).
\\
\\
In Table.\ref{table1} below the energies of the chosen $2l2l'$ levels (where the level $1s$ is considered as a reference) and their different relaxation rates are presented. Collisional rates ($C$) and dielectronic-capture rates ($DC$) are calculated for the temperature $T_e = 500 \ \text{eV}$.

\begin{center}
\begin{tabular}{|c||p{2cm}||c||p{2cm}||c||p{2cm}|}
\hline
Atomic levels: j & Energies [\text{eV}]: $E_j - E_{1s}$ &  {$\Gamma \ [\text{s}^{-1}]$} & $DC$ [$\text{cm}^{3}\text{s}^{-1}$] $T_{e} = 500$ \text{eV}
 & $\gamma^{r} \ [\text{s}^{-1}]$ & $\gamma^{C}$ [$\text{cm}^{3}\text{s}^{-1}$] $T_{e} = 500$ \text{eV} \\
\hline
 $2p^2$ ${}^1D_2$ & $1218$ & $2.74\text{E+}14$ & $8.91\text{E-}13$& $2.91\text{E+}13$ & $5.69\text{E-}09$ \\
\hline
$2s2p$ ${}^1P_1$ & $1220$ & $1.42\text{E+}14$ & $2.75\text{E-}13$& $1.49\text{E+}13$ & $1.33\text{E-}08$ \\
\hline
$2p^2$ ${}^1S_0$ & $1238$ & $1.41\text{E+}13$ & $8.78\text{E-}15$& $2.47\text{E+}13$ & $9.67\text{E-}09$ \\
\hline
\end{tabular}

\captionof{table}{Atomic properties of the three chosen $2l2l'$ energy levels of helium-like aluminium ($Z = 13$). Calculations were performed by the use of the code $\text{FAC}$ \cite{Gu1}.} 
\label{table1}
\end{center}

Electronic inelastic collisional rates between the three energy levels $2p^2$ ${}^1D_2$, $2s2p$ ${}^1P_1$ and $2p^2$ ${}^1S_0$ are given in Table.\ref{table2}. 

\begin{center}
\begin{tabular}{|c||c||c|}
\hline
Atomic energy level: $l$ & Atomic energy level: $u$ &  $C_{l \rightarrow u}$ [$\text{cm}^{3}\text{s}^{-1}$] \\
\hline
 $2p^2$ ${}^1D_2$ &$2s2p$ ${}^1P_1$ & $5.46\text{E-}09$  \\
\hline
$2p^2$ ${}^1D_2$ & $2p^2$ ${}^1S_0$ & $5.31\text{E-}11$  \\
\hline
$2s2p$ ${}^1P_1$ & $2p^2$ ${}^1S_0$ & $2.99\text{E-}09$  \\
\hline
\end{tabular}

\captionof{table}{Inelastic electronic collisional excitation rates for transitions occurring between the three chosen $2l2l'$ energy levels of helium-like aluminium ($Z = 13$). Calculations were performed by the use of the code $\text{FAC}$ \cite{Gu1}.} 
\label{table2}
\end{center}

The evaluation of the plasma ion electric microfield dependent rate Eq.\ref{W_E_gammaJ_gammaJ} is given by inserting the normal value $F_0$ Eq.\ref{F0_Normal_Value} of the Holtsmark \cite{Holstmark1} static distribution of the plasma ion electric microfield. $F_0$ is given by:
\\
\begin{equation}\label{F0_Normal_Value}
F_{0} = 2.6 \ e \ n_{i}^{\frac{2}{3}} \ Z_{i}
\end{equation}
\\
where in Eq.\ref{F0_Normal_Value}, $e$ is the absolute value of the electronic charge, $n_i$ is the ion density and $Z_i = Z - n_b$ is the ionic effective charge defined as the difference between the nuclear charge $Z$ of the ion and the number of bound electrons $n_b$. For numerical calculations we use in Eq.\ref{F0_Normal_Value} the following relation between the electronic density $n_e$ and the ionic density $n_i$:  
\\
\begin{equation}\label{ne_ni}
n_i = \frac{n_e}{Z_i}
\end{equation}
\\
So, we normalize the electric field amplitude $E$ by $F_0$: 
\\
\begin{equation}\label{}
E = F_0 \times \frac{E}{F_0} \equiv F_0 \times \theta
\end{equation}
\\
where we have introduced the dimensionless factor $\theta$ equal to $E$ divided by $F_0$. Then the rate Eq.\ref{W_E_gammaJ_gammaJ} reads:
\\
\begin{equation}\label{W_Ne}
W_{\gamma J , \  \gamma' J'}(E) \equiv W_{\gamma J , \  \gamma' J'}(n_{e}, \theta)
\end{equation}
\\
and:
\\
\begin{multline}\label{W_E_gammaJ_gammaJ_Ne}
W_{\gamma J , \  \gamma' J'}(n_e, \theta) = 218.733 \times \frac{\left(-1\right)^{J-J'} \times C^{\textbf{J'11},\ 00}_{J}}{g_{\gamma J}} \times \frac{g_{\gamma J} \times f_{\gamma J \rightarrow \gamma' J'}}{E_{\gamma' J'} - E_{\gamma J}} \times n_{e}^{\frac{4}{3}} \times Z_{i}^{\frac{2}{3}} \times \theta^{2} \times \\
\frac{\left[ \sum_{\gamma''J''} W_{\gamma J,\ \gamma'' J''} +  \sum_{\gamma''J''} W_{\gamma' J',\ \gamma''J''} \right]}{\left[ \frac{E_{\gamma J}-E_{\gamma' J'}}{\hbar} \right]^2 +\frac{1}{4}\left[ \sum_{\gamma''J''} W_{\gamma J,\ \gamma'' J''} +  \sum_{\gamma''J''} W_{\gamma' J',\ \gamma'' J''} \right]^2}
\end{multline}
\\
where in Eq.\ref{W_E_gammaJ_gammaJ_Ne}, $f_{\gamma J \rightarrow \gamma' J'}$ is the dimensionless oscillator strength for the transition $\gamma J \rightarrow \gamma' J'$ \cite{Cowan1}. In Eq.\ref{W_E_gammaJ_gammaJ_Ne}, the numerical result is in $\text{s}^{-1}$, the energies must be inserted in $\text{eV}$, the electronic density $n_e$ in $\text{cm}^{-3}$ and atomic transition rates in $\text{s}^{-1}$. Below in Table.\ref{table3}, absorption oscillator strengths ($f_{j'  \rightarrow j}$) multiplied by statistical weights of lower energy levels ($j'$) for transitions occurring between the three $2l2l'$ chosen energy levels are given by the use of the code $\text{FAC}$ \cite{Gu1}. The statistical weight of the energy level $2p^2$ ${}^1D_2$ is equal to 5, for the energy level $2s2p$ ${}^1P_1$ is equal to 3 and for the energy level $2p^2$ ${}^1S_0$ is equal to 1.

\begin{center}
\begin{tabular}{|c||c||c|}
\hline
Atomic energy level: $j$ & Atomic energy level: $j'$ &  $g_{j'} \times f_{j' \rightarrow j}$ \\
\hline
 $2s2p$ ${}^1P_1$ & $2p^2$ ${}^1D_2$ & $3.058\text{E-}02$  \\
\hline
$2p^2$ ${}^1S_0$ & $2s2p$ ${}^1P_1$ & $1.591\text{E-}01$  \\
\hline
\end{tabular}

\captionof{table}{Absorption oscillator strength multiplied by lower level statistical weight for transitions occurring between the three $2l2l'$ energy levels of helium-like aluminium (Z = 13). Calculations were carried out with the code $\text{FAC}$ \cite{Gu1}.} 
\label{table3}
\end{center}

In the the following in Table.\ref{table4}  we give numerical values of coefficients $C^{\textbf{J'11},\ 00}_{J}$ Eq.\ref{C1100_gammaJ_delta} associated to this three $2l2l'$ energy levels in cases where the transition oscillator strength is not vanishing $f_{\gamma J \rightarrow \gamma' J'} \neq 0$ Table.\ref{table3}.

\begin{center}
\begin{tabular}{|c||c||c|}
\hline
Atomic energy level: $\gamma J$ & Atomic energy level: $\gamma' J'$ &  $C^{\textbf{J'11},\ 00}_{J}$ \\
\hline
 $2s2p$ ${}^1P_1$ \ ; \ J = 1 & $2p^2$ ${}^1D_2$ \ \ ; \ J' = 2 & $-0.333$  \\
\hline
$2p^2$ ${}^1S_0$ \ \ ; \ J = 0 & $2s2p$ ${}^1P_1$ \ ; \ J' = 1 & $-0.333$  \\
\hline
\end{tabular}

\captionof{table}{Coefficients $C^{\textbf{J'11},\ 00}_{J}$ Eq.\ref{C1100_gammaJ} between the three chosen $2l2l'$ energy levels for transitions where the oscillator strength is not vanishing $f_{\gamma J \rightarrow \gamma' J'} \neq 0$ Table.\ref{table3}.} 
\label{table4}
\end{center}

From Tables.\ref{table3},\ref{table4}, we see that the field dependent rate $W_{\gamma J, \gamma' J'}(E)$ is not vanishing only between the two pairs of energy levels: ($2s2p$ ${}^1P_1$; $2p^2$ ${}^1D_2$) and ($2p^2$ ${}^1S_0$; $2s2p$ ${}^1P_1$). Below in Table.\ref{table5} we give a comparison between values of total collisional-radiative relaxations of the three $2l2l'$ energy levels of helium-like aluminium with their total rates due ti the plasma ion electric microfield mixing dynamics Eq.\ref{W_E_gammaJ_gammaJ_Ne} for one value of the parameter $\theta = 1$ and different values of the electronic density $n_e$ for the electronic temperature $T_e = 500 \ \text{eV}$. The total collisional-radiative relaxation includes: radiative decay ($A$), autoionization ($\Gamma$) and the relaxation by the electronic inelastic collision ($C$).

\begin{center}
\begin{tabular}{|c||c|c||c|c|}
\hline
Electronic density $\rightarrow$ & \multicolumn{2}{ c || }{$10^{+20} \ [\text{cm}^{-3}]$ } &  \multicolumn{2}{ c | }{$10^{+22} \ [\text{cm}^{-3}]$} \\
\hline
Atomic energy levels $\downarrow$ & $W_{CR}$ & W($\theta$, $n_e$) & $W_{CR}$ & W($\theta$, $n_e$) \\
\hline
$2p^2$ ${}^1D_2$ & 3.043E+14 & 1.173E+10 & 3.607E+14 & 7.162E+12  \\
\hline
$2s2p$ ${}^1P_1$ & 1.582E+14 & 1.961E+10 & 2.899E+14 & 1.200E+13  \\
\hline
$2p^2$ ${}^1S_0$ & 3.977E+13 & 2.084E+08 & 1.355E+14 & 1.954E+11\\
\hline
\end{tabular}

\begin{tabular}{|c||c|c||c|c|}
\hline
Electronic density $\rightarrow$ & \multicolumn{2}{ c || }{$10^{+23} \ [\text{cm}^{-3}]$ } &  \multicolumn{2}{ c | }{$10^{+24} \ [\text{cm}^{-3}]$} \\
\hline
Atomic energy levels $\downarrow$ & $W_{CR}$ & W($\theta$, $n_e$) & $W_{CR}$ & W($\theta$, $n_e$)  \\
\hline
$2p^2$ ${}^1D_2$ & 8.728E+14 & 4.824E+14  & 5.994E+15 & 9.176E+15 \\
\hline
$2s2p$ ${}^1P_1$ & 1.487E+15 & 8.119E+14  & 1.346E+16 & 1.661E+16 \\
\hline
$2p^2$ ${}^1S_0$ & 1.006E+15 & 2.388E+13  & 9.709E+15 & 3.940E+15 \\
\hline
\end{tabular}

\captionof{table}{Comparison between total collisional-radiative relaxations of the three $2l2l'$ energy levels of helium-like aluminium with their total rates due to the plasma ion electric microfield lixing dynamics Eq.\ref{W_E_gammaJ_gammaJ_Ne} for one value of the parameter $\theta = 1$, different values of the electronic density $n_e$ and one value of the electronic temperature $T_e = 500 \ \text{eV}$. Rates are given in $\text{s}^{-1}$.} 
\label{table5}
\end{center}

Below in Table.\ref{table6}, the same comparison is given for $\theta = 1.5$.

\begin{center}
\begin{tabular}{|c||c|c||c|c|}
\hline
Electronic density $\rightarrow$ & \multicolumn{2}{ c || }{$10^{+20} \ [\text{cm}^{-3}]$ } &  \multicolumn{2}{ c | }{$10^{+22} \ [\text{cm}^{-3}]$} \\
\hline
Atomic energy levels $\downarrow$ & $W_{CR}$ & W($\theta$, $n_e$) & $W_{CR}$ & W($\theta$, $n_e$) \\
\hline
$2p^2$ ${}^1D_2$ & 3.043E+14 & 2.638E+10 & 3.607E+14 & 1.611E+13  \\
\hline
$2s2p$ ${}^1P_1$ & 1.582E+14 & 4.413E+10 & 2.899E+14 & 2.700E+13  \\
\hline
$2p^2$ ${}^1S_0$ & 3.977E+13 & 4.689E+08 & 1.355E+14 & 4.398E+11\\
\hline
\end{tabular}

\begin{tabular}{|c||c|c||c|c|}
\hline
Electronic density $\rightarrow$ & \multicolumn{2}{ c || }{$10^{+23} \ [\text{cm}^{-3}]$ } &  \multicolumn{2}{ c | }{$10^{+24} \ [\text{cm}^{-3}]$} \\
\hline
Atomic energy levels $\downarrow$ & $W_{CR}$ & W($\theta$, $n_e$) & $W_{CR}$ & W($\theta$, $n_e$)  \\
\hline
$2p^2$ ${}^1D_2$ & 8.728E+14 & 1.085E+15  & 5.994E+15 & 2.065E+16 \\
\hline
$2s2p$ ${}^1P_1$ & 1.487E+15 & 1.827E+15  & 1.346E+16 & 3.736E+16 \\
\hline
$2p^2$ ${}^1S_0$ & 1.006E+15 & 5.373E+13  & 9.709E+15 & 8.864E+15 \\
\hline
\end{tabular}

\captionof{table}{Comparison between total collisional-radiative relaxations of the three $2l2l'$ energy levels of helium-like aluminium with their total rates due to the plasma ion electric microfield mixing dynamics Eq.\ref{W_E_gammaJ_gammaJ_Ne} for one value of the parameter $\theta = 1.5$, different values of the electronic density $n_e$ and one value of the electronic temperature $T_e = 500 \ \text{eV}$. Rates are given in $\text{s}^{-1}$.} 
\label{table6}
\end{center}

From Tables.\ref{table5},\ref{table6}, it is shown that at the electronic temperature $T_e =500 \ \text{eV}$ and at the electronic density $n_e = 10^{+23} \ \text{cm}^{-3}$ the mixing dynamics rate Eq.\ref{W_E_gammaJ_gammaJ_Ne} is of the same order of magnitude than usual collisional-radiative relaxation rates. At larger densities ($n_e > 10^{+23} \ \text{cm}^{-3}$) this rate is even larger. This shows the importance of this rate in high density plasma regimes to diagnose exotic state of matter as well as the evolution of a solid, under X-ray free electron lasers (XFEL's) irradiation, to warm dense matter (WDM) to dense strongly coupled plasma (DSCP).  


\section{Summary}
\hspace{2mm} 

In this paper we presented numerical estimations of the plasma ion electric microfield mixing dynamics rate introduced in \cite{Youcef7}. Comparisons to usual collisional-radiative rates were carried out in the case of three atomic energy levels of the doubly excited $2l2l'$ configuration of helium-like aluminium ($Z = 13$): $2p^2$ ${}^1D_2$, $2s2p$ ${}^1P_1$ and $2p^2$ ${}^1S_0$. The numerical result shows that at high densities this rate is at the same order of magnitude as usual collisional-radiative rates and even exceed them for certain plasma parameters. This demonstrates the potential role of this rate for a better understanding of the heating mechanism underlying the evolution of a solid sate matter to a plasma in the context a solid irradiated by an X-ray free electron laser (XFEL). 
     


\end{document}